\documentclass[journal]{IEEEtran}
\usepackage{color} 
\usepackage[normalem]{ulem}
\usepackage{amsmath, amsthm, amssymb}
\usepackage{epsfig}
\usepackage{latexsym}
\usepackage{graphicx} 
\usepackage{relsize}
\usepackage{cite}
\usepackage{subfigure}
\usepackage{bm}
\usepackage{algorithm}
\usepackage{algpseudocode}
\usepackage{pifont}
\usepackage[utf8]{inputenc}
\usepackage[hidelinks]{hyperref}
\usepackage[english]{babel}
\usepackage{mathtools}
\DeclareGraphicsExtensions{.eps}
\newcommand{\hmisink}{h_{\scriptscriptstyle M_m, FC}}

\newcommand{\gmi}{\gamma_{\scriptscriptstyle M_m}}

\newcommand{\ttot}{T_{\scriptscriptstyle Total}}

\newcommand{\pfcik}{P_{\scriptstyle f}(m,k)}
\newcommand{\pdcik}{P_{\scriptstyle d}(m,k)}

\newcommand{\pfcmssr}{P_{\scriptstyle f}(m,k)}
\newcommand{\pfclssr}{P_{\scriptstyle f}(m,k)}

\newcommand{\emk}{M^{(k)}}
\newcommand{\elk}{L^{(k)}}
\newcommand{\elkopt}{L^{(k)}_{\scriptstyle opt}}

\newcommand{\pfk}{G_f^{\scriptscriptstyle SSR}(k)}
\newcommand{\pdk}{G_d^{\scriptscriptstyle SSR}(k)}

\newcommand{\pfkccs}{G_f^{\scriptscriptstyle CCS}(k)}
\newcommand{\pdkccs}{G_d^{\scriptscriptstyle CCS}(k)}

\title{Sensing-Throughput Tradeoff for Superior Selective Reporting-based Spectrum Sensing in Energy Harvesting HCRNs}

\author{Rajalekshmi Kishore,~\IEEEmembership{Student Member,~IEEE}, Sanjeev~Gurugopinath,~\IEEEmembership{Member, IEEE},\\Sami Muhaidat,  \IEEEmembership{Senior Member,~IEEE}, Paschalis C.~Sofotasios,~\IEEEmembership{Senior Member,~IEEE},\\Octavia~A.~Dobre,~\IEEEmembership{Senior Member, IEEE}, and Naofal Al-Dhahir,~\IEEEmembership{Fellow,~IEEE}

	\thanks{This work is submitted in part to International Conference on Communications (ICC) 2019.}

	\thanks{R. Kishore is with the Department of Electrical and Electronics Engineering, BITS Pilani, K.~K.~Birla Goa Campus, Goa 403726, India (email: {\rm lekshminair2k@yahoo.com}).}

	\thanks{S. Gurugopinath is with the Department of Electronics and Communication Engineering, PES University, Bengaluru 560085, India, (email: {\rm sanjeevg@pes.edu}).}

	\thanks{S.  Muhaidat and P. C. Sofotasios are with the Department of Electrical and Computer Engineering, Khalifa University of Science and Technology, Abu Dhabi, UAE  (emails: {\rm \{muhaidat, p.sofotasios\}@ieee.org}).}
	
	\thanks{Octavia~A.~Dobre is with the Engineering and Applied Science, Memorial University, St. Johns, NL A1B 3X5, Canada (email: {\rm odobre@mun.ca}).}

	\thanks{N. Al-Dhahir is with the Department of Electrical Engineering, University of Texas at Dallas, TX 75080 Dallas, USA (e-mail: {\rm aldhahir@utdallas.edu}).}	
}
 
\begin{document}

 \maketitle
 
\begin{abstract}
In this paper, we investigate the performance of conventional cooperative sensing (CCS) and superior selective reporting (SSR)-based cooperative sensing in an energy harvesting-enabled  heterogeneous cognitive radio network (HCRN). In particular, we derive  expressions for the achievable  throughput of  both schemes and formulate nonlinear integer programming problems, in order to find the throughput-optimal set of spectrum sensors scheduled to sense a particular channel, given  primary user (PU) interference and energy harvesting constraints. Furthermore, we present  novel solutions for the underlying optimization problems based on the cross-entropy (CE) method, and compare the performance  with  exhaustive search and  greedy algorithms. Finally, we discuss the tradeoff between the average achievable throughput of the SSR and CCS schemes, and highlight the regime where the SSR scheme outperforms the CCS scheme. Notably, we show that there is an inherent tradeoff  between the channel available time and the detection accuracy. Our numerical results show that, as the number of spectrum sensors increases, the channel available time gains a higher priority in an HCRN, as opposed to detection accuracy.
\end{abstract}


\IEEEpeerreviewmaketitle

\section{Introduction}
\label{sec:introduction}
Heterogeneous wireless sensor networks (HWSN) are envisioned to address the recent dramatic growth of wireless data services (\cite{Xie_IEEE_2012}, \cite{Qiu_Adhoc_2017}). In order to meet the ever-increasing traffic demands and to maintain the sustainability of wireless networks, there have been extensive research efforts on key enabling technologies for spectral- and energy-efficient future wireless networks \cite{Vincent_book_2017}. Considering the scarcity of spectrum and energy resources, achieving the envisioned sustainability and the efficient utilization of resources are considered as  major challenges. A promising solution to address these challenges is to integrate the cognitive radio (CR) technology  \cite{Mitola_IEEE_1999} with HWSN \cite{Muge_Adhocnetwork_2017},
collectively termed as \emph{heterogeneous cognitive radio networks} (HCRN)  \cite{Zhang_IEEEVehTec_2017}.

In an HCRN, the deployed sensors periodically scan a primary user (PU) spectrum to detect the availability of vacant channels, and subsequently  enable data transmission over a secondary network, while guaranteeing a given PU interference level \cite{Moh_IEEEconf_2010}. However, the periodic sensing increases the energy consumption, which is a critical issue in battery operated sensor networks. To realize a green, sustainable and secure HCRN, tradeoff studies among the detection performance, achievable throughput, energy utilization, and security are critical problems that need to be addressed. Towards this end, HCRNs with energy harvesting (EH) spectrum sensors (\cite{Wu_TWC_2017}, \cite{Tan_Adhocneetwork_2015}) are considered, which enhance both spectrum efficiency and energy efficiency (\cite{Bae_IEEETCom_2016}, \cite{Park_IEEETWC_2013}, \cite{Anastas_AdhocNetwork_2009}, \cite{Ercan_IEEETMC_2017}).

In conventional cooperative spectrum sensing (CCS), a tradeoff exists between the sensing accuracy and data transmission duration, called the sensing-throughput tradeoff \cite{Liang_IEEE_2008}. Sensing accuracy -- in terms of probability of detection -- is hence essential to improve the average throughput, which can be achieved by using the optimal fusion rule, namely, the $L$-out-of-$M$ rule \cite{Varshney_IEEE.tran_IT_1989}. However, as the number of sensors increases, the average throughput decreases due to the increase in the reporting overhead, even though the sensing accuracy increases. Therefore, methods to increase the channel available time by reducing the sensing overhead have also received considerable research attention (\cite{Khan_IEEESPL_2010}, \cite{Firoo_Adhocnetwork_2017}). In \cite{Khan_IEEESPL_2010}, reporting secondary users (SUs) were chosen based on the best individual detection performance. User selection based on uncorrelated decisions across SUs was employed in \cite{Cacciapuoti_IEEEJSAC_2012}, where a dedicated error-free channel was assumed for reporting individual sensing results. The best sensor set selection scheme was proposed as a non-cooperative game in \cite{Yuan_TEEETSP_2011}. A disadvantage in these works is that a reduction in the channel available time occurs, due to the need to report all the associated local decisions to the fusion center (FC), which decreases linearly with the number of SUs  \cite{Monemian_IEEEsensor_2016, Ebrahimzadeh_IEEETVT_2015}. To further reduce the sensing overhead and to improve the channel available time for data transmission, a spectrum sensing (SS) strategy known as superior selective reporting (SSR) scheme was proposed in  \cite{Dai_IEEETVT_2015}, which was shown to achieve a larger probability of detection compared to the CCS strategy with the OR fusion rule \cite{Dai_IEEETVT_2015}. In terms of probability of detection, the SSR scheme is indeed inferior in comparison to the optimal CCS scheme which utilizes the Chair-Varshney ($L$-out-of-$M$) fusion rule \cite{Chair_TAES_1986}. However, the decision reporting overhead in the SSR scheme is significantly reduced, since only one selected node reports its decision to a center node/sink. Therefore, the SSR scheme results in a better data transmission time which enhances the achievable network throughput. Hence, a CR system incorporating the SSR scheme with energy harvesting nodes \cite{Bhowmick_IEEE_2017}, \cite{Shah_Adhobnetwork_2016} achieves a major improvement in the channel available time and network  throughput in an HCRN, for a given primary interference constraint.\\
\\
In this paper, we analyze the throughput performance of SSR-based-multi-channel HCRN, and formulate an optimization problem that maximizes the average achievable throughput to find the best sensor-to-channel assignment vector, subject to energy harvesting and interference constraints. To the best of our knowledge, throughput and sensing-throughput tradeoff analysis based on optimal spectrum sensing allocation for multichannel HCRN based on the CE algorithm have not been considered in the literature.

The main contributions of this paper are summarized as follows:
\begin{itemize}
	\item The average achievable throughput of an SSR-based, multi-channel HCRN is analyzed in terms of the channel available time and detection accuracy.
	\item The problem of finding an optimal set of spectrum sensors scheduled for spectrum sensing for each channel such that the average network throughput is maximized, formulated and solved by employing the cross-entropy (CE) algorithm. The advantages of the CE algorithm in contrast to the exhaustive search algorithm and a greedy algorithm are established. The computational complexity of the CE algorithm is discussed in detail.
	\item It is demonstrated that, as the number of sensors increases, the proposed SSR-based scheme outperforms the CCS scheme that employs the $L$-out-of-$M$ rule in terms of average achievable throughput.
	\item A tradeoff between the average achievable throughput of the SSR and CCS schemes is studied, which is the inherent tradeoff between the channel available time and detection accuracy. In other words, we show that as the number of spectrum sensors increases, the channel available time gets a higher priority in a HCRN than the detection accuracy.
\end{itemize}

\section{Related Work}
In this section, the state-of-the-art literature is discussed which can be classified into two categories, namely the sustainable cognitive radio networks and sensor scheduling approaches for spectrum sensing.

\subsection{Sustainable Cognitive Wireless Sensor Networks}
Battery operated wireless sensors in a WSN usually have a short life time, which directly affects the sustainability of the network. Numerous solutions have been proposed in the literature to address  the sustainability of the network by employing efficient data transmission. Wang et al.~\cite{Wang_IEEETSC_2017} proposed a time adaptive schedule algorithm  for data collection from the WSN to the cloud, along with a minimum cost spanning tree-based routing method to reduce the transmission cost. They showed that their proposed method considerably reduces the latency and optimizes the energy consumption, which makes the sensor-cloud pair sustainable. To prolong the network life time, a sustainable WSN has been considered in \cite{Djenouri_IEEESC_2017} from the perspective of energy-aware communication coverage where two types of sensor nodes, namely energy rich nodes and energy limited nodes are deployed. Bedeer et al. \cite{Bedeer_IEEEVT_2015} proposed a novel optimization algorithm to maximize energy efficiency of OFDM based CR systems under channel uncertainties. Simulation results showed that that the proposed algorithm guarantees a minimum QoS for the SU at the expense of deteriorating the energy efficiency. The same authors in \cite{Bedeer_IEEETWC_2014} solve the problem of jointly maximizing the CR system throughput and minimizing its transmit power, subject to constraints on both SU and PUs by adapting problem of OFDM-based cognitive radio (CR) systems.

Throughput-optimal resource allocation policy design for sustainable energy harvesting (EH)-based WSN (EHWSN) was addressed in \cite{Xu_IEEETWC_2015} and \cite{Zhang_IEEEJSAC_2016}. Xu et al.~\cite{Xu_IEEETWC_2015} investigate the utility-optimal data sensing and transmission in an EHWSN, with heterogeneous energy sources such as power grids and utilizing the harvested energy. They also analyzed the tradeoff between the achieved network utility and cost due to the energy utilized from the power grid. Zhang et al.~in  \cite{Zhang_IEEEJSAC_2016} developed an optimization framework to guarantee sensor sustainability in an EH-based CRN (EHCRSN), where parameters such as stochastic energy harvesting, energy consumption, spectrum utilization and spectrum access processes are designed in an optimal way. An aggregate network utility optimization framework based on a Lyapunov cost-based optimization was developed for the design of online energy management, spectrum management and resource allocation. They also demonstrated that the outcome of the work can be used as a guide for designing a practical EHCRN, which guarantees PU protection and sensors sustainability at the same time. However, these existing methods only offer sustainability of network and are unable to effectively ensure the balance between overall performance, reduction in overhead and network resource utilization.

\subsection{Sensor Scheduling}
Energy-aware sensor scheduling in WSNs has also attracted significant research attention. In \cite{Huang_IEEETSG_2013}, the authors proposed a new priority-based traffic scheduling for CR communication on smart grids, considering channel-switch and spectrum sensing errors, and a system utility optimization problem for the considered communication system was formulated. Such scheduling scheme was shown to serve as a new paradigm of a future CR-based smart grid communication network. More recently, in order to avoid a large overhead and delay, smart scheduling of a collaborative sequential sensing-based wide band detection scheme was proposed in \cite{Zhao_IEEETN_2018} to effectively detect the PU activity in a wide frequency band of interest. A sensor selection scheme was proposed in \cite{Khan_IEEE_2010} to find a set of sensors with the best detection performance for cooperative spectrum sensing, which does not require apriori knowledge of the PU SNR.  The throughput of the CR network is optimized in \cite {Liu_IEEEVT_2015}, by scheduling the spectrum sensing activities based on the residual energy of each sensor. Liu \textit{et al}.~in \cite{Liu_IEEEVT_2015} proposed an ant colony-based energy-efficient sensor scheduling algorithm to optimally schedule a set of sensors to achieve the required sensing performance and to increase the overall CR system throughput. It was demonstrated that the proposed algorithm outperforms a greedy algorithm and the genetic algorithm with a lower computational complexity. However, the sensors employed in the above system model are energy-constrained battery powered sensors and not sensors equipped with energy harvesting. These scheduling strategies do not specifically consider the tradeoff between network performance and resource spectrum utilization in a CRWSN. Moreover, the overhead of network resources caused by the cooperative sensing strategies is not accounted for in the existing methods, which is a key factor. Thus, the problem of sensor scheduling in a CRWSN needs to be considered in terms of a collective network utility and efficiency performance.

\subsection{Comparison with Existing Literature}
	The study in \cite{Rajalekshmi_Adhoc_2017} showed that the SSR-based scheme outperforms the CCS scheme in terms of energy efficiency, but not in the context of an HCRN. Additionally, note that in \cite{Dai_IEEETVT_2015}, the SSR scheme was shown to outperform the OR fusion rule in terms of probability of detection, while we compare the performance of the SSR scheme with the $L$-out-of-$M$ rule, in terms of achievable throughput. Further, the spectrum sensor scheduling problem considered in \cite{Zhang_IEEEVehTec_2017} neither considered the SSR scheme, nor the sensing-throughput tradeoff study in terms of probability of detection and achievable throughput. Moreover, \cite{Zhang_IEEEVehTec_2017} did not consider the $L$-out-of-$M$ rule for the performance study.\\

The remainder of this paper is organized as follows. The system model for multi-channel HCRN employing the SSR scheme is presented in Section \ref{SecSysModel}. The spectrum sensor scheduling problem that maximizes the average achievable throughput for the SSR scheme is formulated and studied in Section \ref{SecProbForm}. The results and discussions are presented in Section \ref{SecResults}, and conclusions are provided in Section \ref{SecConc}.

\section{System Model}\label{SecSysModel}

\begin{figure*}
	\centering
	\includegraphics [width=5in]{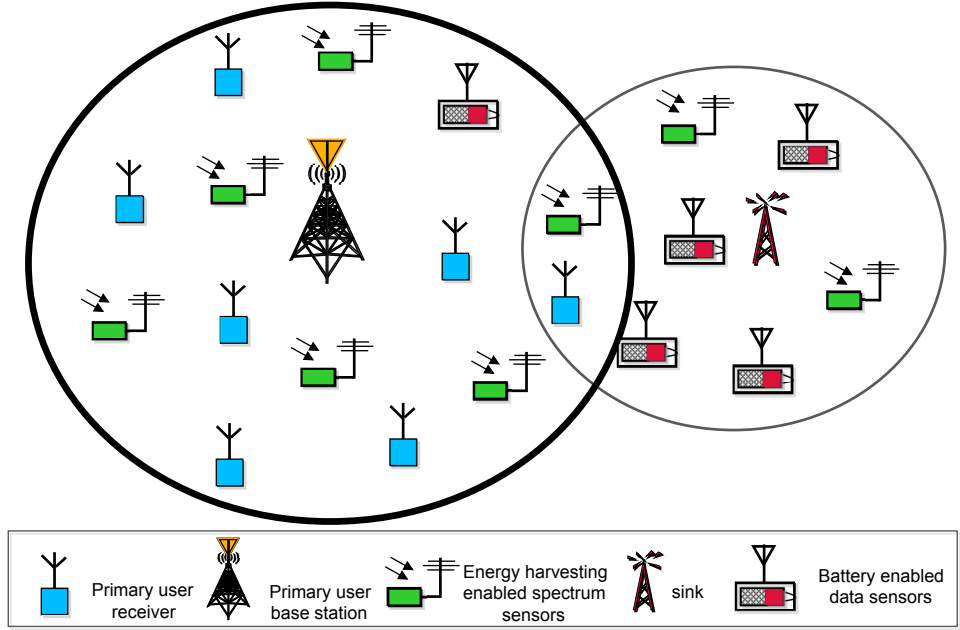}
	\caption{System model of the HCRN.}
	\label{Fig1systemmodel}
\end{figure*}

\begin{figure*}
	\centering
	\includegraphics[scale=0.46]{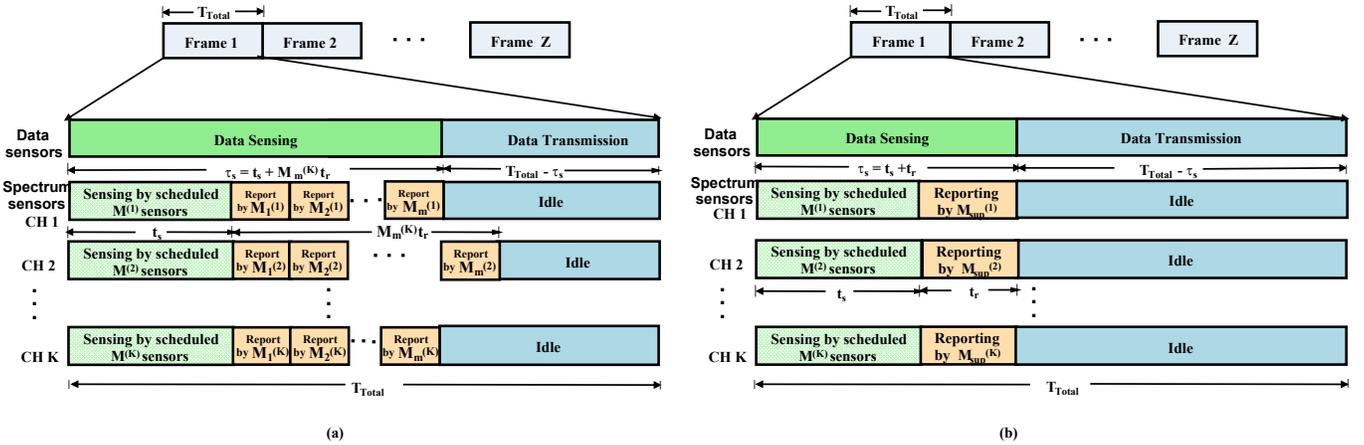}
	\caption{Frame structure of the HCRN for (a) CCS scheme and (b) SSR scheme.}
	\label{Fig1Timeslot}
\end{figure*}

\subsection{Network Architecture} \label{Network}
We consider an HCRN with the following three types of nodes:  EH-enabled spectrum sensors, $N$ battery powered data sensors and a sink (or a fusion center, FC) \cite{Zhang_IEEEVehTec_2017} as shown in Fig.~\ref{Fig1systemmodel}. It is assumed that the PUs are distributed within the coverage area of the HCRN. The licensed spectrum is divided into $K$ non-overlapping channels of equal bandwidth $W$. The data sensors collect data from an area of interest, and transmit it to the sink over licensed channels. It is assumed that there are $K$ transceivers mounted on the sink, such that it can support $K$ concurrent data transmissions over $K$ different non-overlapping channels in each time slot \cite{Zhang_IEEEJSAC_2016}, as shown in the frame structure of the HCRN  in Fig.~\ref{Fig1Timeslot}. Therefore, we assume that each spectrum sensor can sense multiple orthogonal channels simultaneously \cite{Liu_IEEEAccess_2017, Gokceoglu_IEEE_2013}. The availability information of the licensed spectrum is acquired from the EH spectrum sensors. Here, we assume that the spectrum sensors use the power-splitting based energy harvesting  \cite{Huang_CST_2015}. The data sensors utilize the vacant channels declared by the spectrum sensors on a priority basis.\footnote{In this work, we assume that all the sensors faithfully report their decisions to the FC. Analysis on the malicious behavior of spectrum sensors and its impact on the sensing performance is beyond the scope of this work.} The FC controls the scheduling of both the spectrum sensors and data sensors. We consider only the scheduling of the spectrum sensors in this work. The set of spectrum sensors for each channel is assigned using the cross-entropy (CE) algorithm, as discussed in \cite{Zhang_IEEEVehTec_2017}.  For cooperation in sensing, we use the superior selective reporting (SSR) scheme \cite{Dai_IEEETVT_2015} which is explained in the next section. Later, the sink assigns the available channels to the data sensors for data transmission. During the data transmission phase, the data sensors communicate the collected data to the sink. Minimizing the energy consumption of a data sensor is of critical importance since it is assumed to be battery powered. This can be accomplished by optimizing the transmission time and power allocation for the data sensors using a similar setup as described in \cite{Zhang_IEEEVehTec_2017}. However, the optimal scheduling of data sensors, as well as an analysis on the corresponding energy consumption is not considered in this work. On a related note, the setup described in this work can also be considered as a worst-case performance study.

Periodic sensing is carried out with a frame period of $\ttot$ seconds. Each frame duration is divided into two phases, namely a sensing phase and a data transmission phase, with duration given by $\tau_s$ and $\ttot-\tau_s$ seconds, respectively. In the sensing duration $\tau_s$, a preassigned optimal subset of the $M$ spectrum sensors, denoted by $\emk$, $k=1,2,\ldots,K$, simultaneously sense the presence of the PU for a time $t_{s}$, and one among these $\emk$ sensors is selected based on its SNR to report its decision to the sink during reporting time slot $t_{r}$, corresponding to each channel. The advantage of employing the SSR scheme is that it increases the throughput and reduces the sensing overhead when compared to the conventional cooperative sensing (CCS) scheme using the OR rule \cite{Rajalekshmi_Adhoc_2017}. Meanwhile, the data sensors collect information and when the sink identifies all the available channels, the data sensors transmit data by utilizing all the available channels in the data transmission phase for a duration $\ttot-\tau_s$.

\subsection{Conventional Cooperative Sensing (CCS) Scheme} \label{SecCCSScheme}
The CCS scheme is a common technique, where the energy-based sensing is employed during the sensing phase for a duration of $t_s$ seconds where a set of spectrum sensors are assigned to sense the $k^{\text{th}}$ channel. Subsequently, the remaining duration of the sensing time, namely $\tau_s-t_s$, is further divided into $\emk$ sub-slots for the transmission of the individual decisions by the nodes $\{M_{m}^{(k)}, m=1,\ldots, M,~ k=1,\ldots, K\}$ to the sink (FC) (\cite{Althunibat_JCC_2014,Ejaz_wiely_2015,Li_IEEE_2013}). To save on the sensing overhead, it is assumed that each sensor transmits a one-bit decision over a dedicated, error free channel. Therefore, as shown in Fig.~\ref{Fig1Timeslot}(a), the sensing duration adds to a total of $\tau_s=t_s+\emk t_r$ seconds, where $t_r$ denotes the reporting time-duration of each sub-slot. Hence, the sensing time $\tau_s$ increases linearly with $\emk$, which decreases the channel available time and hence the average achievable throughput. At the end of time slot $\tau_s$, the sink collects the sensing results from all the scheduled spectrum sensors and combines these decisions by using a suitable fusion rule such as the AND rule \cite{Nallagonda_WPC_2013}, OR rule \cite{Huang_IEEE_2013} or the $L$-out-of-$M$ rule \cite{Atapattu_IEEE_2011}, and estimates the availability of the channels. In this work, we consider the $L$-out-of-$M$ rule, since it is known to be Bayesian optimal \cite{Varshney_IEEE.tran_IT_1989}. The sensing duration of the CCS scheme increases with $\emk$. To reduce the sensing overhead, a selective reporting based cooperative spectrum sensing scheme, namely the SSR scheme has been proposed \cite{Dai_IEEETVT_2015}, which is briefly explained next.

\subsection{Superior Selective Reporting (SSR)-Based Sensing Scheme} \label{SecSSRScheme}
The SSR scheme, originally proposed in \cite{Dai_IEEETVT_2015}, has multiple advantages over the CCS scheme that employs the OR rule, as the sink receives the decision only from the \emph{superior sensor} denoted by 
\begin{align}
M_{\scriptstyle sup}^{(k)}=\underset{M_{m} \in  \Phi_k}{\arg \max} ~~\bigg( \gmi | \hmisink | ^2\bigg),
\end{align}
where $m=1,\ldots, M$, which is selected based on the received SNR between the FC and sensors, across all sensors.  The set of spectrum sensors $M^{(k)}$ that detect the presence of the PU constitutes a \emph{detection set} $\Phi_k$, $k=1,\ldots,K$. Each sensor $\{M_m \in \Phi_k\}$ sets off a timer at the end of the sensing phase, with each initial value $\{\mathtt{T}_{m}, M_m \in \Phi_k\}$ set inversely proportional to its received SNR $\gmi | \hmisink | ^2$ \cite{Dai_IEEETVT_2015}, where $\gmi$ and $\hmisink$ denote the SNR and the fading coefficient of the channel from the FC to $M_m$, $m=1,\ldots,M$, respectively, i.e., $\mathtt{T}_{m}= \mu/(\gmi \left\vert \hmisink \right\vert^{2})$, for some $\mu \in \mathbb{R}^+$. The sensor with the highest SNR, termed as the superior sensor, exhausts its timer first and reports to the FC. Hence, only the superior sensor sends its local decision to the sink in time slot $t_r$ by transmitting a short duration flag packet, signaling its presence. All other sensors, waiting for their timer to expire, back off immediately as soon as they hear this flag \cite{SINRBletsas_IEEE_2006}. In this work, we assume that all the spectrum sensors are within the PU coverage area, and within the coverage area of each other. Although it is assumed that each spectrum sensor reports only a one-bit decision to the FC, which typically leads to a marginal improvement in overhead, it has been shown that SSR results in a notable improvement in the signal detection, as opposed to the scheme that uses the OR rule \cite{Dai_IEEETVT_2015}. In this work, we further show that the adopted SSR scheme yields a significant improvement in throughput, in comparison with the scheme that employs the $L$-out-of-$M$ rule.

\subsection{Performance Analysis with Energy Detection}
As mentioned earlier, we employ energy detection (ED) in this work to detect the presence of the PU. In this section, we discuss the performance of energy detection strategies that employ the CCS and SSR schemes.

\subsubsection{CCS Scheme}
For the CCS scheme, the probabilities of signal detection and false-alarm at the $m^{\text{th}}$ sensor sensing the $k^{\text{th}}$ channel  are given by \cite{Zhang_IEEEVehTec_2017}
\begin{align}
& \pfcik  =  Q \bigg( \left(  \frac {\varepsilon} {\sigma^{2}}  -   1 \right)  \sqrt{U} \bigg) \triangleq \overline{P_{f}}, \\
&  \pdcik=Q\left ( \frac{Q^{-1}(\overline{P_{f}})-  \sqrt{U}\gamma_{{m}k}}{ \sqrt{2\gamma_{{m}k}+1} } \right ),
\end{align}
where $Q(\cdot)$ is the complementary cumulative distribution function (CDF) of the standard Gaussian distribution, and $\gamma_{{m}k}$ denotes the received SNR from the PU at the $k^{\text{th}}$ channel by the $m^{\text{th}}$ sensor. $U$ is the average number of samples of the received signal at the $m^{\text{th}}$ spectrum sensor on the $k^{\text{th}}$ channel. We assume that the PU signal is a complex-valued PSK signal and the noise is distributed as a circularly symmetric complex Gaussian with zero mean and variance $\sigma^{2}$ \cite{Liang_IEEE_2008}. Without loss of generality, we set the detection threshold $\varepsilon$ to be the same for all the sensors. The overall probabilities of false-alarm and detection at the $k^{\text{th}}$ channel for the CCS scheme are obtained by fixing $ \pfcik$ to a predefined level $\overline{P_{f}} \in (0,1)$, as
\begin{align}  
\label{loutofmPfeqn}
& \pfkccs = \sum_{n=\elk}^{{M^{(k)}}} \binom{{M^{(k)}}}{n} \pfcik(1-\pfcik)^{{{M^{(k)}}-n}} \nonumber \\
& \phantom{\pfkccs} =\sum_{n=\elk}^{{M^{(k)}}}\binom{{M^{(k)}}}{n}\overline{P_{f}}(1-\overline{P_{f}})^{{{M^{(k)}}-n}} \\
& \pdkccs= \sum_{n=\elk}^{{M^{(k)}}}\binom{{M^{(k)}}}{n}\pdcik(1-\pdcik)^{{{M^{(k)}}-n}},
\end{align}
where the total number of cooperating sensors for sensing the $k^{\text{th}}$ channel is  $\emk$, and the value of $L$ determines the fusion rule used. The optimum value of $L$ is given by \cite{Varshney_IEEE.tran_IT_1989}
\begin{align} \label{LoptEqn}
&\hspace{-0.2cm}\elkopt = \min \left ( M^{(k)}, \right. \nonumber \\
& ~~~~~~~~~~~~~ \left. \left \lceil \frac{\log \left(\frac{P(\mathcal{H}_{0})}{1-P(\mathcal{H}_{0})} \right) + M^{(k)} \log \left(\frac{1-{P_{f}(m,k)}}{P_m(m,k)} \right)}{\log\left \{ \left ( \frac{1-P_m(m,k)}{P_f(m,k)} \right )\left ( \frac{1-P_f(m,k)}{P_m(m,k)} \right ) \right \}} \right \rceil \right ),
\end{align}
where only those $P_f(m,k)$ and $P_m(m,k)$ values for $m \in \emk$ are used to evaluate \eqref{LoptEqn} for each $k = 1, \ldots, K$. If $\elk$ is chosen as either $M^{(k)}$, $1$ or $\lceil M^{(k)}/2 \rceil$, the $L$-out-of-$M$ rule reduces to the AND, OR or Majority fusion rules, respectively. As mentioned previously, we mainly consider the optimum fusion rule with $L$ as given in \eqref{LoptEqn}. However, for a comparative study, we consider the CCS scheme with AND and OR rules later, which have their associated advantages and disadvantages \cite{Huang_IEEE_2013,Maleki_IEEE_2011}.

\subsubsection{SSR Scheme} \label{SSRselection}
We follow the method of choosing the superior SU and calculating the received SNR as described in \cite{Dai_IEEETVT_2015, Rajalekshmi_Adhoc_2017}. The probabilities of false-alarm, $\pfk$, and signal detection, $\pdk$, at the  FC are given, respectively, as \cite{Dai_IEEETVT_2015}
\begin{align}
&  \pfk  = \hspace{-0.1cm} \sum_{j=1}^{{2^{M^{(k)}}}-1} \hspace{-0.15cm} \bigg[ \hspace{-0.05cm} \prod_{m \in \Phi_{j,k}} \hspace{-0.2cm} \pfclssr \hspace{-0.2cm} \prod_{m \in  \overline{\Phi}_{j,k}  } \hspace{-0.2cm} (1-\pfcmssr)\bigg] \hspace{-0.15cm} \\
& \phantom{\pfk}  =  1 \hspace{-0.05cm} - \hspace{-0.05cm} (1 \hspace{-0.05cm} - \hspace{-0.05cm} \overline{P_{f}})^{M^{(k)}}, \\
& \pdk  = 1 \hspace{-0.05cm} - \hspace{-0.05cm}\prod_{m =1  }^{M^{(k)}} (1 \hspace{-0.05cm} - \hspace{-0.05cm} \pdcik)^{M^{(k)}}.
\end{align}
Here $\Phi_{j,k}$ is the $j^{\text{th}}$ nonempty sub-collection of detection set $\Phi_k$, and $\overline{\Phi}_{j,k}$ is the complement of $\Phi_{j,k}$. In contrast to the optimal CCS scheme with $L$-out-of-$M$ fusion rule, the advantage of the SSR scheme is in saving the reporting time, which increases the channel available time for data transmission -- vide Fig.~\ref{Fig1Timeslot}, hence,  improving the average achievable throughput for secondary transmission over the $k^{\text{th}}$ channel. Next, we present the main contribution of this paper, i.e., we formulate an optimization problem for finding the best subset of spectrum sensors per channel, denoted by $M^{(k)}$, to maximize the network throughput for a given PU interference constraint.

\section{Problem Formulation: Optimal Scheduling} \label{SecProbForm}
The average number of bits transmitted by the data sensors across all $K$ channels in one time duration is defined as the average achievable throughput of an HCRN \cite{Zhang_IEEEVehTec_2017}. Consider a sensor-to-channel assignment matrix $\mathbf{J} \in \{0,1\}^{M \times K}$. Let the $(m,k)^{\text{th}}$ element $[\mathbf{J}]_{m,k}$, $m=1,\ldots,M$, $k=1,\ldots,K$ of $1$ indicate that the sensor $m$ is scheduled for spectrum sensing for channel $k$, and $0$ otherwise. Our aim is to find the optimal $\mathbf{J}$ that maximizes the average throughput in the considered HCRN.
%\subsection{Achievable Throughput for Multi-Channel Sensing}
The average achievable throughput depends on the available time for data transmission, probability that favors the inactive state of PU, $P(\mathcal{H}_{0})^{(k)}$, of the $k^{\text{th}}$ channel, $\pfcik$, $\pdcik$, and the channel capacity, $\mathtt{C}$. We model the PU dynamics over each channel as a stationary exponential ON-OFF random process \cite{Zhang_IEEEVehTec_2017}, with the average available time of the $k^{\text{th}}$ channel being the product of stay-over time and the stationary state probability. Let $T_{ON}^{(k)} =1/\lambda_{0}^{(k)}$ and $T_{OFF}^{(k)} =1/\lambda_{1}^{(k)}$ be the average values of the stay-over time of the ON state and OFF state of the $k^{\text{th}}$ channel respectively, where $\lambda_{0}^{(k)}$ denotes the transition rate from the ON state to the OFF state on the $k^{\text{th}}$ channel and $\lambda_{1}^{(k)}$ denotes the transition rate in the opposite direction.  The stationary probabilities of the ON and OFF states of the PU on each channel are given by \cite{Zhang_IEEEVehTec_2017}
\begin{equation}
P(\mathcal{H}_{1})^{(k)}=\frac{\lambda_{1}^{(k)}}{\lambda _{1}^{(k)}+\lambda _{0}^{(k)}},  ~~~~  P(\mathcal{H}_{0})^{(k)} =\frac{\lambda _{0}^{(k)}}{\lambda _{1}^{(k)}+\lambda _{0}^{(k)}}.
\end{equation}
The average achievable network throughput under four possible scenarios are as listed as below.  

\textbf{S1:} In this scenario, the spectrum sensors successfully detect the absence of PUs  with probability $P(\mathcal{H}_{0}) ^{(k)}$ $(1-\pfk)$. The throughput for this scenario is expressed as
\begin{align}
& P(\mathcal{H}_{0}) ^{(k)}\left [ 1-\bar{P}_{f} \right ]^{\sum_{m=1}^{M}[J]_{m,k}}~I_{d,SSR}^{(k)}~\mathtt{C}^{(k)}(\ttot-\tau_{s}),
\end{align}
where $I_{d,SSR}^{(k)}$ is a binary variable introduced as a constraint to satisfy the PU protection requirement, defined as  
\begin{equation}
I_{d,SSR}^{(k)} = \begin{cases}1 & \text{if } 1-\pdk<\overline{PM}_{thr},\\ 0 & \text{otherwise.} \end{cases}
\end{equation}  
Similarly, the throughput for the CCS case can be obtained for this scenario (Table \ref{TputTable}) via the corresponding indicator function defined as:
\begin{equation}
I_{d,CCS}^{(k)} = \begin{cases}1 & \text{if } 1-\pdkccs~<~\overline{PM}_{thr},\\ 0 & \text{otherwise.} \end{cases}
\end{equation}
That is, in both cases, if the probability of miss of the $k^{th}$ channel exceeds a predefined threshold $\overline{PM_{thr}} \in (0,1)$, the decision is said to be unreliable for communication over the $k^{th}$ channel.

\textbf{S2:} Here, the sensors correctly detect the PU as active, with probability $P(\mathcal{H}_{1}) ^{(k)} \pdk$, which results in no throughput. Similarly, no throughput can be achieved in the CCS case.

\textbf{S3:} In this scenario, the sensors falsely detect the PU to be present, with probability $P(\mathcal{H}_{0})^{(k)} \pfk$. Here, since the CR network misses a transmission opportunity, the throughput achieved is given by 
\begin{align}
& P(\mathcal{H}_{0}) ^{(k)}\left [1-(1-\overline{P}_{f}) ^{\sum_{m=1}^{M}[\mathbf{J}]_{m,k}}\right] \nonumber \\
& ~~~~~~~~~~~~~~~~~~I_{d,SSR}^{(k)}~\mathtt{C}^{(k)}~(\ttot-\tau_{s})(-\phi),
\end{align}
where $\phi \in (0,1)$ is a suitably chosen penalty factor \cite{Zheng_springer_2017}. Note that a penalty term is introduced in this case to take into account that the CR network missed a transmission opportunity. For simplicity, $\phi$ may as well be chosen to be zero.

\textbf{S4:} In this scenario, the sensors make an incorrect decision that the PU is absent, with probability $P(\mathcal{H}_{1})^{(k)} (1-\pdk)$. The network causes interference to the PU, with a partial throughput of $\kappa$ $P(\mathcal{H}_{1})^{(k)}\left [1-P_{d}(m,k) \right ]^{\sum_{m=1}^{M}[\mathbf{J}]_{m,k}} I_{d,SSR}^{(k)}~\mathtt{C}^{(k)}$ $(\ttot-\tau_{s})$, with some $\kappa \in (0,1)$. Note that a value of any $\kappa \neq 0$ indicates that even though the CR network causes interference to the PU network, it still communicates with a non-trivial data rate. For simplicity, $\kappa$ can be chosen to be zero.

The throughput achieved due to the CCS and SSR schemes across all scenarios are listed in Table \ref{TputTable}, which is shown on top of the next page. Following these cases, the average achievable throughput of the SSR scheme is given by:
\begin{align}\label{objSSS} 
& R_{SSR} =\sum_{k=1}^K  \hspace{-0.1cm} \left\{  P(\mathcal{H}_{0}) ^{(k)} \hspace{-0.1cm} \left [ 1 \hspace{-0.1cm} - \hspace{-0.1cm} \overline{P}_{f} \right ]^{\overset{M}{\underset{m=1}{\sum}} \left [ \mathbf{J} \right ]_{m,k}} \hspace{-0.2cm} - \phi P(\mathcal{H}_{0})^{(k)} \right. \nonumber \\
& ~~~ \left. \left[1 \hspace{-0.1cm} - \hspace{-0.1cm} (1 \hspace{-0.1cm} - \hspace{-0.1cm} \overline{P}_{f})^{ \overset{M}{\underset{m=1}{\sum}} \left [ \mathbf{J} \right ]_{m,k}}\right] + \hspace{-0.1cm} P(\mathcal{H}_{1})^{(k)}  \right. \nonumber \\
& ~~~ \left. \left [ 1 \hspace{-0.1cm} - \hspace{-0.1cm} \pdcik \right ]^{ \overset{M}{\underset{m=1}{\sum}} \left [ \mathbf{J} \right ]_{m,k}} \kappa \right\} I_{d,SSR}^{(k)}~\mathtt{C}^{(k)}(\ttot \hspace{-0.1cm} - \hspace{-0.1cm} \tau_{s}), \hspace{-0.15cm}
\end{align}
for some $0 \leq \kappa < 1 $ and $\phi \geq 0$. On the other hand, the average achievable throughput for the CCS scheme from Table \ref{TputTable} is given by:
\begin{align}
& R_{CCS} = \sum_{k=1}^K  \hspace{-0.1cm} \left\{ \hspace{-0.1cm} P(\mathcal{H}_{0}) ^{(k)} \hspace{-0.1cm} \left\{ \hspace{-0.1cm} 1 \hspace{-0.1cm} - \hspace{-0.2cm} \sum_{n=\elk}^{{M^{(k)}}}\hspace{-0.2cm} \binom{{M^{(k)}}}{n}\overline{P_{f}}(1- \hspace{-0.1cm} \overline{P_{f}})^{{{M^{(k)}}-n}} \hspace{-0.1cm} \right\} \right. \nonumber \\
& ~ \hspace{-0.1cm} \left. - \phi P(\mathcal{H}_{0})^{(k)} \hspace{-0.15cm} \left\{\hspace{-0.1cm} \sum_{n=\elk}^{{M^{(k)}}} \hspace{-0.2cm} \binom{{M^{(k)}}}{n}\overline{P_{f}}(1 \hspace{-0.1cm} - \hspace{-0.1cm} \overline{P_{f}})^{{{M^{(k)}} \hspace{-0.05cm} - \hspace{-0.05cm} n}} \hspace{-0.1cm} \right\} \hspace{-0.1cm} + \hspace{-0.1cm} \kappa P(\mathcal{H}_{1})^{(k)}  \right. \nonumber \\
& ~~~~ \left. \hspace{-0.1cm}  \left\{1  \hspace{-0.1cm} -  \hspace{-0.2cm} \sum_{n=\elk}^{{M^{(k)}}} \hspace{-0.2cm} \binom{{M^{(k)}}}{n}\pdcik(1 \hspace{-0.1cm} - \hspace{-0.1cm} \pdcik)^{{{M^{(k)}}-n}}\right\}  \right\} \nonumber \\
& ~~~~~~~~ \times I_{d,CCS}^{(k)}~\mathtt{C}^{(k)}(\ttot \hspace{-0.1cm} - \hspace{-0.1cm} t_s-{M^{(k)}}t_r). \hspace{-0.15cm} \nonumber \\
& \phantom{R_{CCS}} =\sum_{k=1}^K  \hspace{-0.1cm}   \left\{P(\mathcal{H}_{0}) ^{(k)} \hspace{-0.1cm} (1 \hspace{-0.1cm} - \hspace{-0.1cm} \pfkccs) \hspace{-0.1cm} -  \hspace{-0.1cm} \phi P(\mathcal{H}_{0})^{(k)} \right. \nonumber \\
& ~~~~~~~~~~~~~ \left. \pfkccs + \hspace{-0.1cm} \kappa~ P(\mathcal{H}_{1})^{(k)}  (1-\pdkccs)  \right\} \nonumber \\
& ~~~~~~~~~~~~~~~~~  I_{d,CCS}^{(k)}~\mathtt{C}^{(k)}(\ttot \hspace{-0.1cm} - \hspace{-0.1cm} t_s-{M^{(k)}}t_r). \hspace{-0.15cm} \label{ReqdEqn}
\end{align}
\begin{table*}[t] \tiny
      \centering
         \caption{Throughput achieved for different scenarios using CCS and SSR schemes.}
      \label{TputTable}
        
        \scalebox{0.99}{ \begin{tabular}{ccccccc}
      \hline
      \multicolumn{3}{c}{CCS Scheme}                                                                                                                       &   \multicolumn{3}{c}{SSR Scheme}                                                                                                                                                                                                                                         \\ \hline
      Scenario              &                                    \multicolumn{1}{l}{\begin{tabular}[c]{@{}l@{}}Throughput\\ (bitz/Hz)\end{tabular}} &  & Scenario                                                                                       &                                                                                                              &  \begin{tabular}[c]{@{}l@{}}Throughput\\ (bitz/Hz)\end{tabular} \\ \hline
      $P(\mathcal{H}_{0}) ^{(k)}  (1-\pfkccs)$     &  $I_{d,CCS}^{(k)}~\mathtt{C}^{(k)}(\ttot-(t_s+\emk t_r))$                                                                                  &  & \begin{tabular}[c]{@{}l@{}} $P(\mathcal{H}_{0}) ^{(k)}  (1-\pfk)$ \end{tabular} & \begin{tabular}[c]{@{}l@{}}\end{tabular} & $I_{d,SSR}^{(k)}~\mathtt{C}^{(k)}(\ttot-(t_s+ t_r)$                                      \\ \hline
      $P(\mathcal{H}_{1}) ^{(k)} \pdkccs$               & {0}                                                                                  &  & \begin{tabular}[c]{@{}l@{}}$P(\mathcal{H}_{1}) ^{(k)} \pdk$\end{tabular}  & 
      \begin{tabular}[c]{@{}l@{}}\end{tabular}                                                                                                                            & {0} 
                                           \\ \hline
                                           \\
      $P(\mathcal{H}_{0})^{(k)} \pfkccs$ &  {$I_{d,CCS}^{(k)}~\mathtt{C}^{(k)}~(\ttot-(t_s+\emk t_r))(-\phi)$}                                                                                  &  & $P(\mathcal{H}_{0})^{(k)} \pfk$                                                                            &                                            \begin{tabular}[c]{@{}l@{}}  \end{tabular}                                                                                 & $I_{d,SSR}^{(k)}~\mathtt{C}^{(k)}~(\ttot-(t_s+ t_r))(-\phi)$                                      \\ \hline
      \\
     $P(\mathcal{H}_{1})^{(k)} (1-\pdkccs)$ & \multicolumn{1}{l}{$I_{d,CCS}^{(k)}~\mathtt{C}^{(k)}$ $(\ttot-(t_s+\emk t_r                        ))(\kappa)$}            &  & $P(\mathcal{H}_{1})^{(k)} (1-\pdk)$                                                                            &                                            \begin{tabular}[c]{@{}l@{}} \end{tabular}                                                                                   & $I_{d,SSR}^{(k)}~\mathtt{C}^{(k)}$ $(\ttot-(t_s+ t_r))(\kappa)$ \\ \hline 
       %$*where$ $0<\kappa_c <1$    
       \end{tabular} } 
      \end{table*}

For the spectrum sensor scheduling problem, we set constraints related to the EH dynamics to facilitate the sustainability of the sensors. In a given frame $\ttot$, the energy consumption of each sensor should not exceed the EH rate, i.e., $\big( \sum_{k=1}^{K} [\mathbf{J}]_{m,k} \big)e_{s} \leq   \delta _{m}~\ttot ~\forall m$, where $\delta_{m}$ is the EH rate. Now, the problem to find the optimum $\mathbf{J}$ that maximizes $R_{SSR}$ can be formulated as follows:
\begin{align}
\mathcal{OP}_{\scriptscriptstyle SSR}: \max_{\mathbf{J}}&\quad R_{SSR}\\ \text{s.t.}& \hspace{-0.2cm} \quad \begin{cases}\big( \sum_{k=1}^{K} [\mathbf{J}]_{m,k} \big)e_s \leq  \delta _{m}~\ttot, ~~\forall\, m\\ [\mathbf{J}]_{m,k}=\{\text{0},\text{1}\}, ~~~\forall m, k\end{cases} \hspace{-0.4cm}\nonumber
\end{align} 
Similarly, the throughput optimization problem governing the CCS scheme is given by
\begin{align}
\mathcal{OP}_{\scriptscriptstyle CCS}: \max_{\mathbf{J}}&\quad R_{CCS}\\ \text{s.t.}& \hspace{-0.2cm} \quad \begin{cases}\big( \sum_{k=1}^{K} [\mathbf{J}]_{m,k} \big)e_s \leq  \delta _{m}~\ttot, ~~\forall\, m\\ [\mathbf{J}]_{m,k}=\{\text{0},\text{1}\}, ~~~\forall m, k\end{cases} \hspace{-0.4cm}\nonumber
\end{align} 
From \eqref{objSSS}, it is clear that as more channels are assigned to a given set of sensors, i.e., as $ \sum_{k=1}^{K} [\mathbf{J}]_{m,k}$ increases, the value of $(1-\overline{P}_{f})^{\sum_{m=1}^{M}[\mathbf{J}]_{m,k}}$ decreases, and $I_{d,SSR}$ tends to unity. Therefore, there is a tradeoff between the values of $(1-\bar{P_{f}})^{\sum_{m=1}^{M}[\mathbf{J}]_{m,k}}$ and $I_{d,SSR}$. As a consequence, as $M$ increases, there exist a tradeoff between the detection accuracy and the channel available time, which affects the average achievable throughput of the network. The $\mathcal{OP}_{\scriptscriptstyle CCS}$ and $\mathcal{OP}_{\scriptscriptstyle SSR}$ are integer programming problems that can be solved by using an exhaustive search method. However, this leads to a search space of $2^{MK}$ elements, which is computationally expensive. Hence, we apply the CE algorithm, as discussed in \cite{Zhang_IEEEVehTec_2017}. Towards this end, the problem  $\mathcal{OP}_{\scriptscriptstyle SSR}$ is transformed into the following unconstrained optimization problem, by applying a penalty of $\omega \in \mathbb{R}^+$ for violating any of the constraints \cite{Zhang_IEEEVehTec_2017}:
\begin{align} 
\max_{\mathbf{J}} ~~~ R_{SSR} -\omega I_{ \hspace{-0.1cm}\left(\overset{K}{\underset{k=1}{\sum}} [\mathbf{J}]_{m,k} e_{s}>\delta_{m}\ttot \right)~, \hspace{-0.1cm}} \label{UnconOPEqn}
\end{align}
The unconstrained optimization problem for the CCS case can be written as
\begin{align} 
\max_{\mathbf{J}} ~~~ R_{CCS} -\omega I_{ \hspace{-0.1cm}\left(\overset{K}{\underset{k=1}{\sum}} [\mathbf{J}]_{m,k} e_{s}>\delta_{m}\ttot \right) , \hspace{-0.1cm}} \label{UnconOPEqnccs}
\end{align}
where $I(\cdot)$ is the indicator function. When the solution violates the constraints, the objective function evaluates to a negative value, which is discarded. In the next section, we discuss the utility of the CE algorithm to solve the above problem, with a discussion on its advantages and computational complexity.

\section{The Cross-Entropy Algorithm}
The CE algorithm is implemented as discussed next \cite{Zhang_IEEEVehTec_2017}. Initially, the iteration counter is set as $i=1 \text{ to } i_{\max} \in \mathbb{Z}^+$. Let $\mathbf{C}$ be the set of all possible $K$-dimensional binary vectors, with $|\mathbf{C}|=2^K$. To begin with, the row vectors of $\mathbf{J}$ are drawn from the matrix $\mathbf{C}$. Now, $Z$ samples of the channel matrix, defined as $\mathbf{V}^{(z)} = \mathbf{v}^{(z)}_{m,\mathbf{c}}, 1 \leq m \leq M, \mathbf{c} \in \mathbf{C},$ $z = 1,\ldots, Z$ of size $M \times 2^K$. Here, $\mathbf{v}^{(z)}_{m,\mathbf{c}}$ denotes the $\mathbf{c}^{\text{th}}$ column vector or $\mathbf{V}^{(z)}$. These column vectors are generated based on a probability mass function (PMF) matrix $\mathbf{Q}^{(i)} = \mathbf{q}^{(i)}_{m,\mathbf{c}}, 1\leq m \leq M, \mathbf{c} \in \mathbf{C}$, where $\mathbf{q}^{(i)}_{m,\mathbf{c}}$ denotes the probability vector that the sensor $m$ is scheduled to sense the channel $k$ in vector $\mathbf{C}$. Now, we calculate the cost function in \eqref{UnconOPEqn} for each sample $z$, and arrange them in descending order. We retain $0 \leq \rho \leq 1$ fraction of the sorted objective function in $\mathcal{OP}_{\scriptscriptstyle SSR}^{(z)}$ and discard all other values. Let the smallest chosen value of the objective function be $\eta$, corresponding to the index $\lceil\rho Z\rceil$. In each step, the PMF matrix is updated as $\mathbf{q}_{m,\mathbf{c}}^{(i+1)}=\frac{\sum_{z=1}^{Z} \textbf{v}^{(z)}_{m,\mathbf{c}} I_{(O^{z}\ge\eta)}}{\lceil\rho Z\rceil}$. The algorithm is stopped either after $i_{\max}$ iterations, or if the stopping criterion $\epsilon > 0$ is satisfied. The resultant $\mathbf{V}^{(z)}$ is selected to map the solution, i.e., the optimal $\mathbf{J}$. To summarize, each iteration of the CE algorithm consists of the steps described in Algorithm~\ref{CEalgo}. A similar procedure is carried out to evaluate the optimal $\mathbf{J}$ for the CCS scheme.

\begin{algorithm}
	\caption{Cross-entropy (CE) algorithm}	\label{CEalgo}
	\begin{algorithmic}[1]
		\Procedure{Initialization}{}\\
		Step 1: \For {i $\leftarrow$ 1 to $i_{\max}$}
		\State $\mathbf{q}^{(1)}_{m,\mathbf{c}}  = 1/|\mathbf{C}|=1/2^K$
		\For {z $\leftarrow$ 1 to Z}\\
		Step 2:   	
		Generate Z samples of matix $\mathbf{V}^{(z)}$ based on PMF matix $\mathbf{Q}^{(i)} = \mathbf{q}^{(i)}_{m,\mathbf{c}}$
		\EndFor \\
		Step 3: \For {z $\leftarrow$ 1 to Z}  \\
		~~~~~~~ Calculate the Objective function in \eqref{UnconOPEqn} $\mathcal{OP}_{\scriptscriptstyle SSR}^{(z)}$
		\EndFor \\
		Step 4:
		~~~~~~~~~ Arrange $\{\mathcal{OP}_{\scriptscriptstyle SSR}^{(z)}, z=1,\ldots,Z\}$ in the decreasing order \\
		Step 5:
		\State Retain $0 \leq \rho \leq 1$ fraction of sorted values $\{\mathcal{OP}_{\scriptscriptstyle SSR}^{(z)}\}$ and discard others. 
		\State Let the smallest chosen value of $\mathcal{OP}_{\scriptscriptstyle SSR}^{(z)}$ be $\eta$, corresponding to the index $\lceil\rho Z\rceil$.\\
		Step 6:
		
		\For {j $\leftarrow$ 1 to M}  
		\For { $~\mathbf{c}~=~1:\mathbf{C}~} ~~~$ Update $\mathbf{q}_{m,c}^{i+1}$ using
		\State  $\mathbf{q}_{m,\mathbf{c}}^{(i+1)}=\frac{\sum_{z=1}^{Z} \textbf{v}^{(z)}_{m,\mathbf{c}} I_{(O^{z}\ge\eta)}}{\lceil\rho Z\rceil}$.
		\EndFor
		\EndFor
		\EndFor \\	   	
		Step 7:	\State Return $\mathbf{V}^{(z)}$, if $i=i_{\max}$ or  \\
		Step 8: \\~~~~~~~~  The channels-to-sensors assignment in $\mathbf{V}^{(z)}$ is mapped to the channels-to-sensors assignment in $\mathbf{J}$ which is a solution to the original optimization problem $\mathcal{OP}_{\scriptscriptstyle SSR}^{(z)}$.
		
		\EndProcedure
	\end{algorithmic}
\end{algorithm}

\subsection{Convergence and Optimality}
The performance of the CE algorithm mainly depends on the speed of convergence and the quality of the obtained solution. The convergence and optimality of the CE algorithm has been previously studied for a variety of combinatorial optimization problems, which mainly involves updating the underlying probability mass function -- given in step $6$ in Algorithm 1. The goal is to eventually converge to a PMF that generates samples close to optimal value of channel assignment matrix $\mathbf{V}^{(z)}$, with high probability. The convergence of CE optimization is not guaranteed in general, but the algorithm is usually found to be convergent for several combinatorial optimization problems of practical relevance \cite{CEalgo_springer_2014}. For most combinatorial problems of interest, the CE algorithm provably converges with probability $1$ to a unit mass density, which always generates samples equal to a single point \cite{Andre_ORletter_2007,Busoniu_IEEEconf_2009}.

The optimality and quantification of performance bounds of the CE algorithm remains an open theoretical issue \cite{CEalgo_springer_2014}. However, in our problem, the number of iterations required for the algorithm to converge depends on the parameters $\rho$ and $\epsilon$. Furthermore, as will be discussed in Sec.~\ref{SecResults}, the convergence of the algorithm can be ensured to be arbitrarily close to the optimal solution at the expense of a larger number of iterations, and a stringent stopping criterion. That is, the probability that the CE algorithm converges to an optimal solution can be made arbitrarily close to $1$, at the expense of convergence time. Moreover, convergence to an optimal solution can be further ensured by using adaptive smoothing techniques \cite{Andre_ORletter_2007,Busoniu_IEEEconf_2009}.

\subsection{Computational Complexity}
In this section, we discuss the computational complexity of the CE algorithm, using an approach discussed in \cite{CEalgo_springer_2014}. The computational complexity of the algorithm, as seen from Algorithm \ref{CEalgo}, will be quantified in terms of $n \triangleq M2^K$, since the calculations involved will be on the $M \times 2^K$ channel assignment matrix which is computationally equivalent to that of an $n$-dimensional Bernoulli distributed vector. Let us further define

\begin{equation}
\kappa_{n} \triangleq i_{\max, n}(Z_{n} Q_{n} + U_{n}),
\end{equation}

where $\kappa_n$ quantifies the total computational complexity of the CE algorithm, $i_{\max, n}$ is the total number of iterations needed before the CE algorithm is stopped, $Z_{n}$ is the sample size of channel assignment matrix $V^{(z)}$, which is generated based on the Bernoulli PMF in each iteration, $Q_{n}$ is the cost of generating a random Bernoulli vector of size $n$,   $U_{n}$ is the combination of the computational cost in updating both the objective function $\mathcal{OP}_{\scriptscriptstyle SSR}^{(z)}$ and the  channel vector assignment probability $\mathbf{q}^{(i)}_{m,\mathbf{c}}$. 

From our simulations, we found that the complexity of $i_{\max, n} = \mathcal{O}(\ln{n})$, for moderately large $n$. The cost of generating a random Bernoulli vector of size $n$ is $Q(n)=\mathcal{O}(n)$. The computations required to select the best $\lceil\rho Z_{n}\rceil$ points from the sample population is given by $\mathcal{O}(\rho Z_{n})$. The combined cost of updating the objective function, sorting the sample population in ascending order and updating the PMF is given as $U(n)= \mathcal{O}(n^3)$. Hence, the overall computational complexity of the CE algorithm for the proposed sensor scheduling problem is given by $\kappa_{n}=\mathcal{O}(n^3 \ln n)$.

\section{Results and Discussion}\label{SecResults}
In this section, we discuss the performance of SSR-based sensing scheme in HCRN in terms of average achievable throughput, and compare its performance with the CCS scheme following the $L$-out-of-$M$ rule, with an optimum $L$ chosen as in \cite{Varshney_IEEE.tran_IT_1989}. Unless otherwise stated, the values of the parameters used are chosen from \cite{Pei_IEEEJSAC_2011}, \cite{Zhang_IEEEVehTec_2017}, and are listed in Table \ref{table}. The sensors are randomly placed in a circular area where the primary user coexists. The channel gain from the PU transmitter to the sensor is calculated as $1/D^\alpha$, where D is the distance between the PU and the spectrum sensors and $\alpha$ is the path-loss exponent. The achievable rates by the data sensors are chosen to be $C=\log_2(1+SNR)=6.658$ bits/sec/Hz \cite{Zhang_IEEEVehTec_2017}. 

\begin{table}
\centering
\caption{Parameter Settings}
\label{table}
\setlength{\tabcolsep}{10pt}
\begin{tabular}
	{|p{125pt}|p{80pt}|}
\hline
			\textbf{Parameters}                                                         & \textbf{Settings }                 \\ \hline
		
			Number of spectrum sensors $M$                                       & 10                          \\
			Number of data sensors $N$                                           & 30                          \\
			Target false alarm probability  $\bar{P_f}$                        & 0.1                         \\
			Target miss- detection probability $\bar{P_m}$                     & 0.1                         \\
			Number of licensed channels                                         & 7                           \\
			Bandwidth of the licensed channel $W$                                & 6 MHz                        \\
			Path-loss exponent $\alpha$                                        & 3.5                         \\
			Transition rate of PU from ON state to OFF state  $\lambda_{0}^{k}$ & 0.6,0.8,1,1.2,1.4,1.6,1.8    \\
			Transition rate of PU from OFF state to ON state $\lambda_{1}^{k}$  & 0.4,0.8,0.6,1.6,1.2,1.4,1.8  \\
			Total frame length  $\ttot$                                    & 100 ms                     \\
			Sampling rates of spectrum sensors U                               & 6000                        \\
			Duration of spectrum sensing phase  $\tau_s$                       & 7 ms                  \\
			Duration of spectrum sensing by assigned sensors on each channel  $t_s$                       & 6 ms                       \\
			Duration of reporting sensing results to sink $t_r$               & 1 ms                       \\
			Sensing power of spectrum sensors  $P_s$                           & 0.1 W                        \\
			Transmission power of data sensors  $P_t$                          & 0.22 W                       \\
			Energy consumption per spectrum sensing                            & 0.11 mJ                      \\
			Fraction of samples retained in CE algorithm  $\rho$               & 0.6                        \\
			Stopping threshold  $\epsilon$                                     & $10^{-3}$                    \\
			partial throughput factor $\kappa$                                 & 0.5                         \\
			Penalty factor for miss detection $\phi$                                              & 0.5                         \\
			SNR of secondary transmission                                      & 20 dB                        \\ \hline
\end{tabular}
\label{tab1}
\end{table}

\begin{figure}[ht]
	\centering
	\includegraphics[scale=0.45]{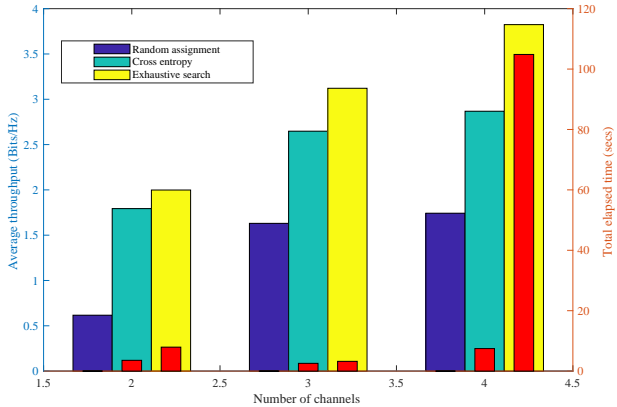}  
	\caption{Average achievable throughput vs.~Number of channels for the SSR-based CE algorithm, random assignment and exhaustive search methods.}
	\label{Fig2thrvschannel}
\end{figure}

\begin{figure}[ht]
\centering
	%	\vspace{-0.5cm}
	\includegraphics [scale=0.6]{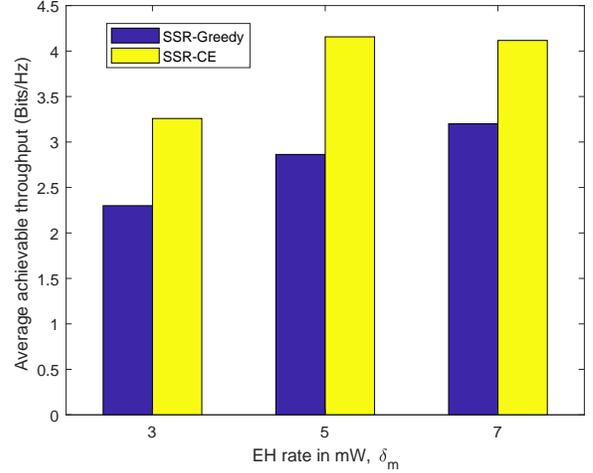} 
	%\vspace{-0.55cm}
	\caption{Comparison of performance of the CE algorithm and the greedy algorithm, for a range of EH rates.}
	\label{fig9}
\end{figure}

The variation of throughput with different number of licensed channels, $K$, is shown in Fig.~\ref{Fig2thrvschannel}. For illustration purposes, we choose $M=3$, and a small $K$, so that a solution using the exhaustive search can be quickly evaluated [5]. Even with small values of $K$, we show that the CE algorithm offers a significant saving in the computation time over the exhaustive search. Moreover, increasing $K$ will not result in a change in the performance trends across all the algorithms. The average achievable throughput of the SSR-based approach using the CE algorithm is compared with the random assignment and exhaustive search methods. The set of all possible assignments is considered in the exhaustive search to find the optimal set, whereas a licensed channel is uniformly and randomly assigned to the spectrum sensors in the random assignment method. As shown in Fig.~\ref{Fig2thrvschannel},  the average achievable throughput obtained by the SSR-based CE algorithm is about $75\%$--$90\%$ of that obtained by the exhaustive search. In contrast, the total elapsed time for the evaluation using the exhaustive search method is about $14$ times longer than that using the CE algorithm, when $K$ is increased to $4$. As $K$ further increases, the elapsed time increases exponentially for the exhaustive search. Thus, the SSR-based CE algorithm attains the maximum throughput with much shorter computation time when compared to the exhaustive search. Figure \ref{fig9} shows the comparison between the performance of the CE algorithm and that of a greedy algorithm \cite{Yu_GLOBECOM_2011}, for different values of EH rates. The greedy algorithm assigns a channel to each sensor sequentially that gives the maximum achievable throughput. It is shown that the CE algorithm outperforms greedy algorithm in terms of the achievable throughput, over a range of EH rates. 

\begin{figure}[ht]
	\centering
	%	\vspace{-0.5cm}
	\includegraphics [scale=0.4]{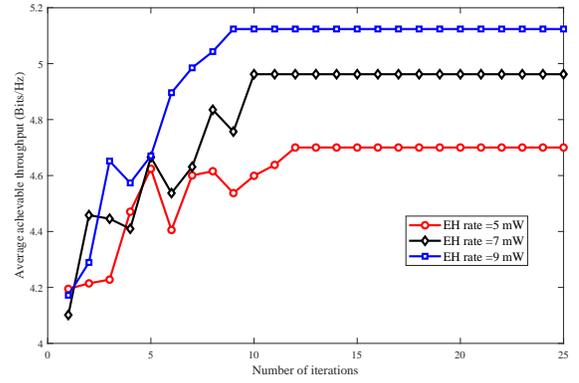} 
	%\vspace{-0.55cm}
	\caption{Average achievable throughput vs.~Number of iteration for different EH rates.}
	\label{FigEHrate}
\end{figure} 

\begin{figure}[ht]
	\centering
	%\vspace{-0.5cm}
	\includegraphics [scale=0.4]{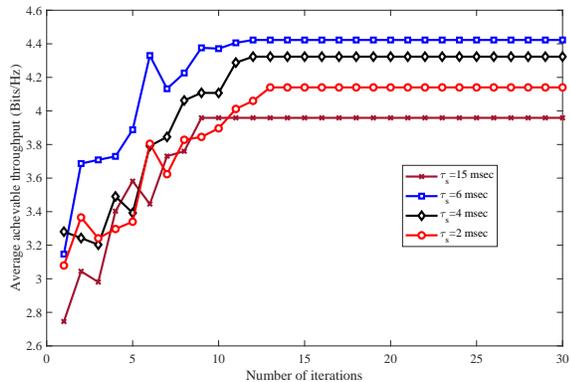} 
	%\vspace{-0.55cm}
	\caption{Average achievable throughput vs.~Number of iteration for different sensing phase durations $\tau_s$.}
	\label{Figtaus}
\end{figure}

The stability of the CE algorithm with respect to the average throughput is shown in Fig.~\ref{FigEHrate}. Here, the convergence of the CE algorithm with the number of iterations can be seen, for different EH rate values. As expected, the average throughput increases with the EH rate. Figure \ref{Figtaus} shows the convergence result of the CE algorithm with respect to the sensing phase duration $\tau_s$ ranging from as low as $2$ ms to a relatively high value such as $15$ ms, for a fixed EH rate of $7$ mW. Note that the achievable throughput first increases with an increase in $\tau_s$ and later decreases as $\tau_s$ is increased further. This concave behavior is due to the sensing-throughput tradeoff \cite{Liang_IEEE_2008}.         

\begin{figure}[ht]
	\centering
	%\vspace{-0.5cm}
	\includegraphics [scale=0.425]{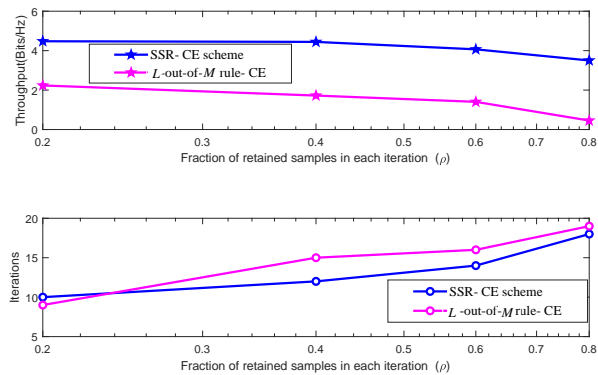} 
	%\vspace{-0.55cm}
	\caption{Impact of the fraction of retained samples $\rho$ on the performance of the CE algorithm}
	\label{fig8}
\end{figure}

Figure \ref{fig8} shows the impact of the fine-tuning CE algorithm parameter, i.e., fraction of samples retained, $\rho$, on the number of iterations and average throughput. It is evident from both plots that CE algorithm with the SSR performs better than that with the CCS with $L$-out-of-$M$ rule. Moreover, the CE algorithm converges quickly with small $\rho$. For the parameters considered in this paper, $\rho$ is chosen to be $0.6$.

\begin{figure}
	\centering
	%	\vspace{-0.5cm}
	\includegraphics [scale=0.4]{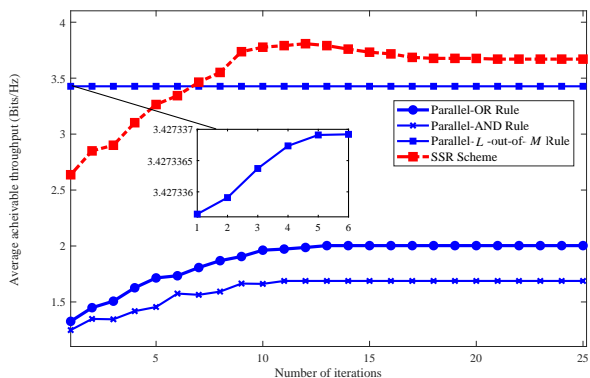} 
	%\vspace{-0.55cm}
		\centering
	\caption{Average throughput vs Number of Iteration.}
	\label{FigAvgthrvsiter}
\end{figure}
\begin{figure}
	%\centering
	%	\vspace{-0.5cm}
	\includegraphics [scale=0.4]{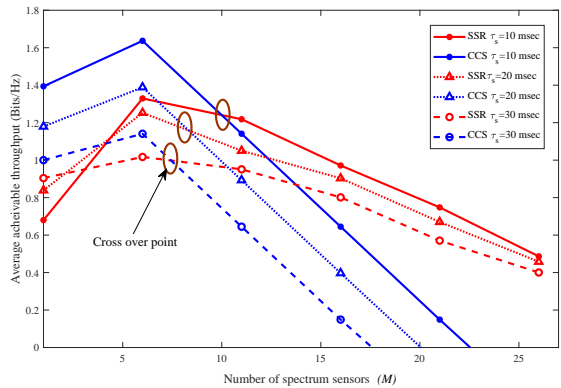} 
	%\vspace{-0.55cm}
	\centering
	\caption{Average achievable throughput vs Number of spectrum sensors, M.}
	\label{ssrkoutmrule}
\end{figure}  

Now, for a network with $M=15$ and $K=7$, the average achievable throughput of the SSR-based CE algorithm is compared with the conventional fusion rules such as OR, AND, and $L$-out-of-$M$ rule, as shown in Fig.~\ref{FigAvgthrvsiter}. In the SSR scheme, since only one sensor reports its decision to the sink, it performs better than the CCS scheme employing $L$-out-of-$M$, OR and AND rules. As expected, the $L$-out-of-$M$ rule performs the best among the CCS schemes, when the optimum value of $L$ is chosen \cite{Varshney_IEEE.tran_IT_1989}. Finally, we discuss the tradeoff between the optimal performance of the SSR-based multichannel scheme with that of the $L$-out-of-$M$ rule based CCS scheme. The variation of average achievable throughput with $M$, for different sensing times $\tau_s$ is shown in Fig.~\ref{ssrkoutmrule}. When $M$ is small, the $L$-out-of-$M$ rule yields a larger throughput due to the better detection accuracy at the expense of relatively less channel available time, as opposed to the SSR scheme which saves the channel available time, but loses out on detection accuracy. Interestingly, as $M$ increases, the SSR scheme outperforms the CCS scheme, as although the detection accuracy of the CCS scheme increases, it loses out on the channel available time. Hence, this tradeoff yields a regime  where SSR is preferred over $L$-out-of-$M$ rule-based CCS scheme. Inherently, this tradeoff is between the detection accuracy and channel available time for secondary data transmission. Therefore, as $M$ increases, the channel available time gets a higher priority as opposed to the detection accuracy in the HCRN, resulting in the SSR scheme as a better choice. However, in the scenario where the detection accuracy is a main concern, the $L$-out-of-$M$ rule can still be employed.

\section{Conclusion and Future Work} \label{SecConc}
We investigated the maximum achievable throughput of SSR-based spectrum sensing in a multichannel HCRN. We quantified the impact of the EH rate on the maximum achievable throughput of the SSR scheme. We have shown that the achievable throughput increases with the EH rate by optimally scheduling the spectrum sensors to sense a particular channel. Through numerical results, we showed that the SSR-based multichannel scheduled sensing scheme outperforms the CCS scheme employing the optimal $L$-out-of-$M$ rule, and discussed the tradeoff between the average achievable throughput of both schemes. We showed that this tradeoff is the inherent tradeoff between the channel available time and the detection accuracy, and discussed the regime where the SSR scheme is preferred over the CCS scheme. The results show that the SSR scheme outperforms the CCS scheme when the number of spectrum sensors is large, and therefore, the channel available time gets a higher priority in an HCRN than the detection accuracy. Hence, in a scenario where spectral efficiency needs to be improved, SSR is a better choice. CCS should be employed in the scenario where the PU protection and detection accuracy are important. As a part of the future work, the optimal power and resource allocation for data sensors is an interesting extension to this problem. 

\bibliographystyle{IEEEtran}                      
\bibliography{IEEEabrv,CEbibloTSC}

\begin{thebibliography}{10}
\providecommand{\url}[1]{#1}
\csname url@samestyle\endcsname
\providecommand{\newblock}{\relax}
\providecommand{\bibinfo}[2]{#2}
\providecommand{\BIBentrySTDinterwordspacing}{\spaceskip=0pt\relax}
\providecommand{\BIBentryALTinterwordstretchfactor}{4}
\providecommand{\BIBentryALTinterwordspacing}{\spaceskip=\fontdimen2\font plus
\BIBentryALTinterwordstretchfactor\fontdimen3\font minus
  \fontdimen4\font\relax}
\providecommand{\BIBforeignlanguage}[2]{{%
\expandafter\ifx\csname l@#1\endcsname\relax
\typeout{** WARNING: IEEEtran.bst: No hyphenation pattern has been}%
\typeout{** loaded for the language `#1'. Using the pattern for}%
\typeout{** the default language instead.}%
\else
\language=\csname l@#1\endcsname
\fi
#2}}
\providecommand{\BIBdecl}{\relax}
\BIBdecl

\bibitem{Xie_IEEE_2012}
R.~Xie, F.~R. Yu, H.~Ji, and Y.~Li, ``Energy-efficient resource allocation for
  heterogeneous cognitive radio networks with femtocells,'' \emph{{IEEE} Trans.
  Wireless Commun.}, vol.~11, no.~11, pp. 3910--3920, Nov. 2012.

\bibitem{Qiu_Adhoc_2017}
T.~Qiu, N.~Chen, K.~Li, D.~Qiao, and Z.~Fu, ``Heterogeneous ad hoc networks:
  Architectures, advances and challenges,'' \emph{Ad Hoc Networks}, vol.~55,
  no. Supplement C, pp. 143 -- 152, Feb. 2017.

\bibitem{Vincent_book_2017}
V.~W.~S. Wong, R.~Schober, D.~W.~K. Ng, and L.-C. Wang, \emph{Key Technologies
  for {5G} Wireless Systems}.\hskip 1em plus 0.5em minus 0.4em\relax Cambridge
  University Press, 2017, 2017.

\bibitem{Mitola_IEEE_1999}
J.~Mitola and G.~Q. Maguire, ``Cognitive radio: making software radios more
  personal,'' \emph{{IEEE} Personal Commun. Mag.}, vol.~6, no.~4, pp. 13--18,
  Aug. 1999.

\bibitem{Muge_Adhocnetwork_2017}
M.~E. Özçevik, B.~Canberk, and T.~Q. Duong, ``End to end delay modeling of
  heterogeneous traffic flows in software defined {5G} networks,'' \emph{Ad Hoc
  Networks}, vol.~60, no. Supplement C, pp. 26 -- 39, May 2017.

\bibitem{Zhang_IEEEVehTec_2017}
D.~Zhang, Z.~Chen, J.~Ren, N.~Zhang, M.~K. Awad, H.~Zhou, and X.~S. Shen,
  ``Energy-harvesting-aided spectrum sensing and data transmission in
  heterogeneous cognitive radio sensor network,'' \emph{IEEE Trans.\ Veh.\
  Technol.}, vol.~66, no.~1, pp. 831--843, Jan. 2017.

\bibitem{Moh_IEEEconf_2010}
M.~I.~B. Shahid and J.~Kamruzzaman, ``Interference protection in cognitive
  radio networks,'' in \emph{Proc. VTC}, May 2010, pp. 1--5.

\bibitem{Wu_TWC_2017}
Q.~Wu, G.~Y. Li, W.~Chen, D.~W.~K. Ng, and R.~Schober, ``An overview of
  sustainable green {5G} networks,'' \emph{IEEE Trans.\ Wireless Commun.},
  vol.~24, no.~4, pp. 72--80, Aug. 2017.

\bibitem{Tan_Adhocneetwork_2015}
Q.~Tan, W.~An, Y.~Han, Y.~Liu, S.~Ci, F.-M. Shao, and H.~Tang, ``Energy
  harvesting aware topology control with power adaptation in wireless sensor
  networks,'' \emph{Ad Hoc Networks}, vol.~27, no. Supplement C, pp. 44 -- 56,
  Apr. 2015.

\bibitem{Bae_IEEETCom_2016}
Y.~H. Bae and J.~W. Baek, ``Achievable throughput analysis of opportunistic
  spectrum access in cognitive radio networks with energy harvesting,''
  \emph{IEEE Trans.\ Commun.}, vol.~64, no.~4, pp. 1399--1410, Apr. 2016.

\bibitem{Park_IEEETWC_2013}
S.~Park, H.~Kim, and D.~Hong, ``Cognitive radio networks with energy
  harvesting,'' \emph{IEEE Trans.\ Wireless Commun.}, vol.~12, no.~3, pp.
  1386--1397, Mar. 2013.

\bibitem{Anastas_AdhocNetwork_2009}
G.~Anastasi, M.~Conti, M.~D. Francesco, and A.~Passarella, ``Energy
  conservation in wireless sensor networks: A survey,'' \emph{Ad Hoc Networks},
  vol.~7, no.~3, pp. 537 -- 568, May 2009.

\bibitem{Ercan_IEEETMC_2017}
A.~Ercan, M.~O. Sunay, and I.~F. Akyildiz, ``{RF} energy harvesting and
  transfer for spectrum sharing cellular {IoT} communications in {5G}
  systems,'' \emph{{IEEE} Trans. Mobile Comput.}, vol.~17, no.~7, pp.
  1680--1694, Jul. 2018.

\bibitem{Liang_IEEE_2008}
Y.~C. Liang, Y.~Zeng, E.~C.~Y. Peh, and A.~T. Hoang, ``Sensing-throughput
  tradeoff for cognitive radio networks,'' \emph{IEEE Trans.\ Wireless
  Commun.}, vol.~7, no.~4, pp. 1326--1337, Apr. 2008.

\bibitem{Varshney_IEEE.tran_IT_1989}
I.~Y. Hoballah and P.~K. Varshney, ``Distributed {B}ayesian signal detection,''
  \emph{{IEEE} Trans. Inf. Theory}, vol.~35, no.~5, pp. 995--1000, Sep. 1989.

\bibitem{Khan_IEEESPL_2010}
Z.~Khan, J.~Lehtomaki, K.~Umebayashi, and J.~Vartiainen, ``On the selection of
  the best detection performance sensors for cognitive radio networks,''
  \emph{IEEE Signal Process.\ Lett.}, vol.~17, no.~4, pp. 359--362, Apr. 2010.

\bibitem{Firoo_Adhocnetwork_2017}
F.~B. Saghezchi, A.~Radwan, and J.~Rodriguez, ``Energy-aware relay selection in
  cooperative wireless networks: An assignment game approach,'' \emph{Ad Hoc
  Networks}, vol.~56, no. Supplement C, pp. 96 -- 108, Mar. 2017.

\bibitem{Cacciapuoti_IEEEJSAC_2012}
A.~S. Cacciapuoti, I.~F. Akyildiz, and L.~Paura, ``Correlation-aware user
  selection for cooperative spectrum sensing in cognitive radio ad hoc
  networks,'' \emph{IEEE {J.} Sel.\ Areas Commun.}, vol.~30, no.~2, pp.
  297--306, Feb. 2012.

\bibitem{Yuan_TEEETSP_2011}
W.~Yuan, H.~Leung, S.~Chen, and W.~Cheng, ``A distributed sensor selection
  mechanism for cooperative spectrum sensing,'' \emph{IEEE Trans.\ Signal
  Process.}, vol.~59, no.~12, pp. 6033--6044, Dec. 2011.

\bibitem{Monemian_IEEEsensor_2016}
M.~Monemian, M.~Mahdavi, and M.~J. Omidi, ``Optimum sensor selection based on
  energy constraints in cooperative spectrum sensing for cognitive radio sensor
  networks,'' \emph{{IEEE} Sensors J.}, vol.~16, no.~6, pp. 1829--1841, Mar.
  2016.

\bibitem{Ebrahimzadeh_IEEETVT_2015}
A.~Ebrahimzadeh, M.~Najimi, S.~M.~H. Andargoli, and A.~Fallahi, ``Sensor
  selection and optimal energy detection threshold for efficient cooperative
  spectrum sensing,'' \emph{{IEEE} Trans. Veh. Technol.}, vol.~64, no.~4, pp.
  1565--1577, Apr. 2015.

\bibitem{Dai_IEEETVT_2015}
Z.~Dai, J.~Liu, and K.~Long, ``Selective-reporting-based cooperative spectrum
  sensing strategies for cognitive radio networks,'' \emph{IEEE Trans.\ Veh.\
  Technol.}, vol.~64, no.~7, pp. 3043--3055, Jul. 2015.

\bibitem{Chair_TAES_1986}
Z.~Chair and P.~K. Varshney, ``Optimal data fusion in multiple sensor detection
  systems,'' \emph{{IEEE} Trans. Aerosp. Electron. Syst.}, vol.~22, no.~1, pp.
  98--101, Jan. 1986.

\bibitem{Bhowmick_IEEE_2017}
A.~Bhowmick, K.~Yadav, S.~D. Roy, and S.~Kundu, ``Throughput of an energy
  harvesting cognitive radio network based on prediction of primary user,''
  \emph{{IEEE} Trans. Veh. Technol.}, vol.~66, no.~9, pp. 8119--8128, Sep.
  2017.

\bibitem{Shah_Adhobnetwork_2016}
S.~T. Shah, K.~W. Choi, S.~F. Hasan, and M.~Y. Chung, ``Throughput analysis of
  two-way relay networks with wireless energy harvesting capabilities,''
  \emph{Ad Hoc Networks}, vol.~53, no. Supplement C, pp. 123 -- 131, Dec. 2016.

\bibitem{Wang_IEEETSC_2017}
T.~Wang, Y.~Li, G.~Wang, J.~Cao, M.~Z.~A. Bhuiyan, and W.~Jia, ``Sustainable
  and efficient data collection from {WSN}s to cloud,'' \emph{IEEE Trans. Sust.
  Comput.}, pp. 1--1, Mar. 2017.

\bibitem{Djenouri_IEEESC_2017}
D.~Djenouri and M.~Bagaa, ``Energy-aware constrained relay node deployment for
  sustainable wireless sensor networks,'' \emph{IEEE Trans. Sust. Comput.},
  vol.~2, no.~1, pp. 30--42, Jan. 2017.

\bibitem{Bedeer_IEEEVT_2015}
E.~Bedeer, O.~Amin, O.~A. Dobre, M.~H. Ahmed, and K.~E. Baddour,
  ``Energy-efficient power loading for {OFDM}-based cognitive radio systems
  with channel uncertainties,'' \emph{{IEEE} Trans. Veh. Technol.}, vol.~64,
  no.~6, pp. 2672--2677, Jun. 2015.

\bibitem{Bedeer_IEEETWC_2014}
E.~Bedeer, O.~A. Dobre, M.~H. Ahmed, and K.~E. Baddour, ``A multiobjective
  optimization approach for optimal link adaptation of {OFDM}-based cognitive
  radio systems with imperfect spectrum sensing,'' \emph{{IEEE} Trans. Wireless
  Commun.}, vol.~13, no.~4, pp. 2339--2351, Apr. 2014.

\bibitem{Xu_IEEETWC_2015}
W.~Xu, Y.~Zhang, Q.~Shi, and X.~Wang, ``Energy management and cross layer
  optimization for wireless sensor network powered by heterogeneous energy
  sources,'' \emph{IEEE Trans.\ Wireless Commun.}, vol.~14, no.~5, pp.
  2814--2826, May 2015.

\bibitem{Zhang_IEEEJSAC_2016}
D.~Zhang, Z.~Chen, M.~K. Awad, N.~Zhang, H.~Zhou, and X.~S. Shen,
  ``Utility-optimal resource management and allocation algorithm for energy
  harvesting cognitive radio sensor networks,'' \emph{IEEE {J.} Sel.\ Areas
  Commun.}, vol.~34, no.~12, pp. 3552--3565, Dec. 2016.

\bibitem{Huang_IEEETSG_2013}
J.~Huang, H.~Wang, Y.~Qian, and C.~Wang, ``Priority-based traffic scheduling
  and utility optimization for cognitive radio communication
  infrastructure-based smart grid,'' \emph{IEEE Trans. Smart Grid}, vol.~4,
  no.~1, pp. 78--86, Mar. 2013.

\bibitem{Zhao_IEEETN_2018}
J.~Zhao, Q.~Liu, X.~Wang, and S.~Mao, ``Scheduling of collaborative sequential
  compressed sensing over wide spectrum band,'' \emph{IEEE/ACM Trans.\
  Networking}, vol.~26, no.~1, pp. 492--505, Feb. 2018.

\bibitem{Khan_IEEE_2010}
Z.~Khan, J.~Lehtomaki, K.~Umebayashi, and J.~Vartiainen, ``On the selection of
  the best detection performance sensors for cognitive radio networks,''
  \emph{IEEE Signal Process.\ Lett.}, vol.~17, no.~4, pp. 359--362, Apr. 2010.

\bibitem{Liu_IEEEVT_2015}
X.~Liu, B.~G. Evans, and K.~Moessner, ``Energy-efficient sensor scheduling
  algorithm in cognitive radio networks employing heterogeneous sensors,''
  \emph{{IEEE} Trans. Veh. Technol.}, vol.~64, no.~3, pp. 1243--1249, Mar.
  2015.

\bibitem{Rajalekshmi_Adhoc_2017}
{R.~Kishore, Ramesha C.~K., S.~Gurugopinath and Anupama K.~R.}, ``Performance
  analysis of superior selective reporting-based energy efficient cooperative
  spectrum sensing in cognitive radio networks,'' \emph{Ad Hoc Networks},
  vol.~65, pp. 99--116, Jul. 2017.

\bibitem{Liu_IEEEAccess_2017}
X.~Liu, F.~Li, and Z.~Na, ``Optimal resource allocation in simultaneous
  cooperative spectrum sensing and energy harvesting for multichannel cognitive
  radio,'' \emph{IEEE Access}, vol.~5, pp. 3801--3812, Mar. 2017.

\bibitem{Gokceoglu_IEEE_2013}
A.~Gokceoglu, S.~Dikmese, M.~Valkama, and M.~Renfors, ``Enhanced energy
  detection for multi-band spectrum sensing under {RF} imperfections,'' in
  \emph{Proc. CROWNCOM}, Jul. 2013, pp. 55--60.

\bibitem{Huang_CST_2015}
X.~Huang, T.~Han, and N.~Ansari, ``On green-energy-powered cognitive radio
  networks,'' \emph{{IEEE} Commun. Surveys Tuts.}, vol.~17, no.~2, pp.
  827--842, Secondquarter 2015.

\bibitem{Althunibat_JCC_2014}
S.~Althunibat, M.~Di~Renzo, and F.~Granelli, ``{Cooperative spectrum sensing
  for cognitive radio networks under limited time constraints},''
  \emph{{Journal on Computer Communications}}, vol.~43, pp. 55--63, May 2014.

\bibitem{Ejaz_wiely_2015}
W.~Ejaz, G.~A. Shah, N.~U. Hasan, and H.~S. Kim, ``Energy and throughput
  efficient cooperative spectrum sensing in cognitive radio sensor networks,''
  \emph{Transactions on Emerging Telecommunications Technologies}, vol.~26,
  no.~7, pp. 1019--1030, 2015.

\bibitem{Li_IEEE_2013}
X.~Li, J.~Cao, Q.~Ji, and Y.~Hei, ``Energy efficient techniques with sensing
  time optimization in cognitive radio networks,'' in \emph{Proc.\ {WCNC}},
  Apr. 2013, pp. 25--28.

\bibitem{Nallagonda_WPC_2013}
S.~Nallagonda, S.~D. Roy, and S.~Kundu, ``Performance evaluation of cooperative
  spectrum sensing scheme with censoring of cognitive radios in rayleigh fading
  channel,'' \emph{Wireless Pers.\ Commun.}, vol.~70, no.~4, pp. 1409--1424,
  Jun. 2013.

\bibitem{Huang_IEEE_2013}
D.~Huang, G.~Kang, B.~Wang, and H.~Tian, ``Energy-efficient spectrum sensing
  strategy in cognitive radio networks,'' \emph{{IEEE} Commun. Lett.}, vol.~17,
  no.~5, pp. 928--931, May 2013.

\bibitem{Atapattu_IEEE_2011}
S.~Atapattu, C.~Tellambura, and H.~Jiang, ``Energy detection based cooperative
  spectrum sensing in cognitive radio networks,'' \emph{IEEE Trans.\ Wireless
  Commun.}, vol.~10, no.~4, pp. 1232--1241, Apr. 2011.

\bibitem{SINRBletsas_IEEE_2006}
A.~Bletsas, A.~Khisti, D.~P. Reed, and A.~Lippman, ``A simple cooperative
  diversity method based on network path selection,'' \emph{IEEE {J.} Sel.\
  Areas Commun.}, vol.~24, no.~3, pp. 659--672, Mar. 2006.

\bibitem{Maleki_IEEE_2011}
S.~Maleki, S.~P. Chepuri, and G.~Leus, ``Energy and throughput efficient
  strategies for cooperative spectrum sensing in cognitive radios,'' in
  \emph{Proc.\ { Signal Processing Advances in Wireless Communications}}, Jun.
  2011, pp. 71--75.

\bibitem{Zheng_springer_2017}
Y.~Zheng and L.~Zheng, ``Sensing transmission tradeoff over penalty for miss
  detection in cognitive radio network,'' \emph{Wireless Personal
  Communications}, vol.~92, no.~3, pp. 1089--1105, Feb. 2017.

\bibitem{CEalgo_springer_2014}
R.~Y. Rubinstein and D.~P. Kroese, \emph{The Cross Entropy Method: {A} Unified
  Approach To Combinatorial Optimization, Monte-carlo Simulation (Information
  Science and Statistics).}\hskip 1em plus 0.5em minus 0.4em\relax Springer,
  2004.

\bibitem{Andre_ORletter_2007}
A.~Costa, O.~D. Jones, and D.~Kroese, ``Convergence properties of the
  cross-entropy method for discrete optimization,'' \emph{Operations Research
  Letters}, vol.~35, no.~5, pp. 573 -- 580, 2007.

\bibitem{Busoniu_IEEEconf_2009}
L.~Busoniu, D.~Ernst, B.~D. Schutter, and R.~Babuska, ``Policy search with
  cross-entropy optimization of basis functions,'' in \emph{2009 IEEE Symposium
  on Adaptive Dynamic Programming and Reinforcement Learning}, Mar. 2009, pp.
  153--160.

\bibitem{Pei_IEEEJSAC_2011}
Y.~Pei, Y.~C. Liang, K.~C. Teh, and K.~H. Li, ``Energy-efficient design of
  sequential channel sensing in cognitive radio networks: Optimal sensing
  strategy, power allocation, and sensing order,'' \emph{IEEE {J.} Sel.\ Areas
  Commun.}, vol.~29, no.~8, pp. 1648--1659, Sep. 2011.

\bibitem{Yu_GLOBECOM_2011}
H.~Yu, W.~Tang, and S.~Li, ``Optimization of cooperative spectrum sensing in
  multiple-channel cognitive radio networks,'' in \emph{Proc.\ GLOBECOM}, Dec.
  2011, pp. 1--5.

\end{thebibliography}

\end{document}